\documentclass[
    ,final            
  ]
  {aipproc}

\layoutstyle{8x11single}

\usepackage{subfig}
\usepackage{graphicx}
\usepackage{wrapfig}

\begin{document}

\title{Recent results and open questions on collective type phenomena from A-A to pp collisions}


\author{M. Petrovici, C. Andrei, I. Berceanu, A. Bercuci, A. Herghelegiu, A. Pop}{
 address={National Institute for Physics and Nuclear Engineering (IFIN-HH), R-077125 Bucharest, 
  Romania}
}

\begin{abstract}
A review of the main results on the collective type expansion of the compressed and hot fireball 
formed in heavy ion collisions and some remarks to be considered when comparing multiplicity wise 
phenomena taking place in A-A, \mbox{p-A} and pp collisions, are followed by a discussion of the
experimental results which seem to evidence collective type phenomena in pp collisions at $\sqrt{s}$ = 7 TeV at 
high charged particle multiplicity. 
Correlations among the kinetic freeze-out temperature, the average transverse expansion velocity and its profile, as a 
function of centrality and multiplicity, extracted from the fits of experimental transverse momentum 
spectra with 
an expression inspired by hydrodynamical models, estimates on Bjorken energy densities and perspectives 
in selecting soft and close to azimuthal isotropic events in pp collisions are presented.

\end{abstract}

\maketitle


\section{1. Introduction}

\begin{figure}[ht]
  \includegraphics[height=.3\textheight]{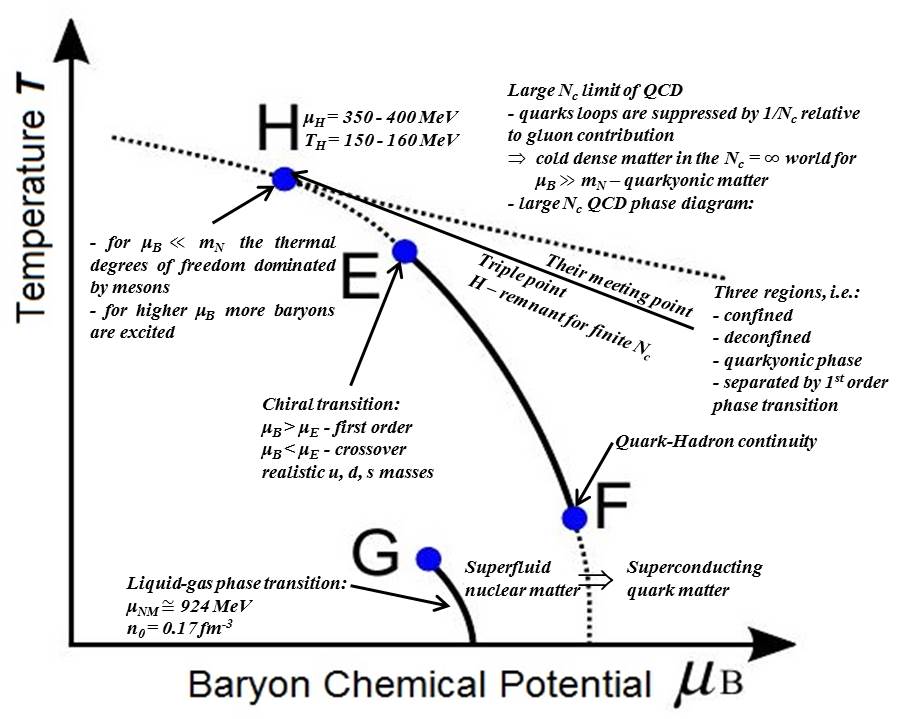}
  \caption{QCD expectations on critical points, triple point and phase transitions of the strongly 
interacting matter phase diagram \cite{fukushima}.}
\label{fig:Fig1}
\end{figure}

   Strongly interacting matter in equilibrium can be characterized by two parameters, i.e. 
temperature T and baryon number density $\rho_{B}$ or its conjugate 
   variable, baryon chemical potential $\mu_{B}$. 
It is well known by now that the basic property of Quantum Chromo-Dynamics (QCD), 
a non-Abelian gauge theory of quarks and gluons, is the asymptotic freedom, i.e. 
the QCD running coupling constant $\alpha_s(Q^2)$ becomes small for a momentum 
transfer Q much larger than the QCD intrinsic parameter $\Lambda$. 
Therefore, for $Q^2\gg \Lambda^2$ a perturbative description of weakly interacting 
quarks and gluons becomes applicable. 
For $Q^2 \sim \Lambda^2$, quarks and gluons are strongly bound in clusters called hadrons.
As \mbox{$\Lambda_{QCD} \sim$ 200 MeV}, a phase transition from hadrons as strongly bound 
clusters of quarks and gluons to a deconfined state of matter formed by weakly interacting 
quarks and gluons is expected at temperature T$\sim\Lambda_{QCD}\sim O(10^{12}K)$ or 
baryon density $\rho_B\sim\Lambda_{QCD}^3\sim1fm^{-3}$. 
In reality, the expectations based on QCD on the phase diagram of strongly interacting 
matter are more complex \cite{fukushima}, as it is shown in Fig.~\ref{fig:Fig1}. 
Besides the local gauge symmetry which is exact in the limit of vanishing quark masses, 
Chiral models predict the existence of a critical point E. 
For realistic u, d and s quark masses the Chiral transition becomes a 
first order transition for $\mu_{B}$ > $\mu_{E}$ and a cross over for $\mu_{B}$ < $\mu_{E}$.
There are also predictions for another critical point F, at low T and high $\mu_B$ ($\mu_F$, $T_F$).
 Below this point, for the cold QCD matter with three degenerate flavors, there is 
no border between superfluid nuclear matter and superconducting quark matter.
The large $N_c$ limit of QCD predicts a suppression of the quark loops relative to the gluon 
contribution \cite{hooft,witten}, an increase in the baryon number density at $\mu_B$ larger 
than the lowest baryon mass $M_B$ being expected.
This cold and dense matter was called quarkyonic matter \cite{larren}.
Therefore, in the large $N_c$ limit of QCD, the phase diagram has three regions, i.e. confined, 
deconfined and quarkyonic, separated by first order phase transitions.  

The temperature T and chemical potential $\mu_B$ can be extracted in the framework of statistical
models, which assume a gas of mesons, baryons and 
resonances in thermal equilibrium, by fitting the yield ratios at different collision energies.
The extracted T and $\mu_B$ clusterize along a curve in the T - $\mu_B$ phase space. 
Along this line, which is not associated to any phase space boundaries, the thermal degrees of 
freedom are dominated by 
mesons for $\mu_{B}  \ll M_{N}$ while at higher $\mu_B$ more baryons are excited. 
This suggests that there is a transitional change at the point ($T_H$, $\mu_H$) where the role of 
baryons in thermodynamics surpasses that of mesons. 
According to statistical models this is located in the region 
of  ~$\mu_H$ = 350 - 400 MeV and $T_H$ = 150 - 160 MeV. 
To what extent such a point is the correspondent of the triple point for 
finite $N_c$ \cite{andronic} is still under debate. 
In nature, according to the Big Bang cosmology, matter composed of quarks and gluons, 
characterized by negligible baryon chemical potential and high temperature, existed 
at a few microseconds after the Big Bang while the core of neutron stars could be 
the place where QCD matter at low temperature could exist.
To what extent such states of matter could be produced and studied in the 
terrestrial laboratories is a fascinating question. 
By colliding heavy ions at relativistic and ultra-relativistic energies, 
transient pieces of matter at densities and temperatures discussed above can be produced. 
However, a few aspects which have to be considered from the very beginning are worth 
to be mentioned, i.e. even in the case of colliding the heaviest nuclei, the size of the 
created system is rather small, its initial state is highly non-homogeneous and dynamical 
effects play a crucial role, the system being characterized by a violent evolution. 
All these considerations have to be taken into account when trying to reconstruct  
detailed information on the dynamics and properties of each step of the evolution following 
the initial phase of the collision, from the particles and electromagnetic radiation reaching the detectors. 
At ultra-relativistic energies, even the hadrons become rather complex objects. 
A free hadron could be considered in each moment as a cloud of quasireal partons belonging to a 
cascade whose density, seen by a 
parton of a similar cascade from the colliding partner, increases with the energy being expected 
to reach a saturation at very high energy \cite{gribov}. 
If a parton of a cascade meets on its way a parton of the colliding partner, interacts with it, 
the whole cascade changes, 
its coherence is broken, the partons cannot assembly back and they continue to live and decay in 
secondary hadrons and last but not least the 
struck cascade could interact with the others. 
Such multiparton interactions and rescatterings could contribute to a large energy transfer 
in a collision volume of proton size, easily reaching the deconfinement conditions. 
If this is the case and considering that the mean free path of the constituents in a deconfined 
medium is of the order of 0.2 - 0.3 fm, a 
close to equilibrium deconfined initial state could be expected in very high energy pp collisions. 
Therefore, quite probable at such energies, a piece of matter of proton size, with a radius of few 
times larger than the mean free path, 
expands hydrodynamically once the energy transfer is significantly large, i.e. low 
impact parameter - high charged particle multiplicity.

A review of the main results on the collective type expansion of the compressed and hot 
fireball formed in heavy ion collisions starting from 
SIS18 energies up to the highest one of $\sqrt{s_{NN}}$ = 2.76 TeV at LHC is presented in Section 2. 
Section 3 is dedicated to some considerations on the differences one has to take
into account in 
comparing multiplicity wise phenomena taking place in A-A, p-A and pp collisions.
Experimental evidence on similar trends in the shape of transverse momentum ($p_{T}$) spectra of 
identified charged hadrons and their 
ratios in Pb-Pb at $\sqrt{s_{NN}}$ = 2.76 TeV and p-Pb at $\sqrt{s_{NN}}$ = 5.02 TeV collisions as a function of centrality and multiplicity classes, 
respectively and in pp collisions at $\sqrt{s}$ = 7 TeV as a function of 
charged particle multiplicity are discussed in Section 4. 
In the same section,  correlations among the kinetic freeze-out temperature, the
average  transverse expansion velocity 
and its profile, as a function of centrality and multiplicity, extracted from the
fits of experimental $p_T$ spectra with an expression inspired by hydrodynamical models are presented.
Estimates on the Bjorken energy density and perspectives in selecting soft and close to azimuthal 
isotropic events in pp collisions are also included in this section.
The conclusions are presented in Section 5.

\section{2. $<\beta_T>$, $T_{kin}$, $T_{ch}$ - collision energy dependence}

\begin{figure}[ht]
\includegraphics[height=0.26\textheight]{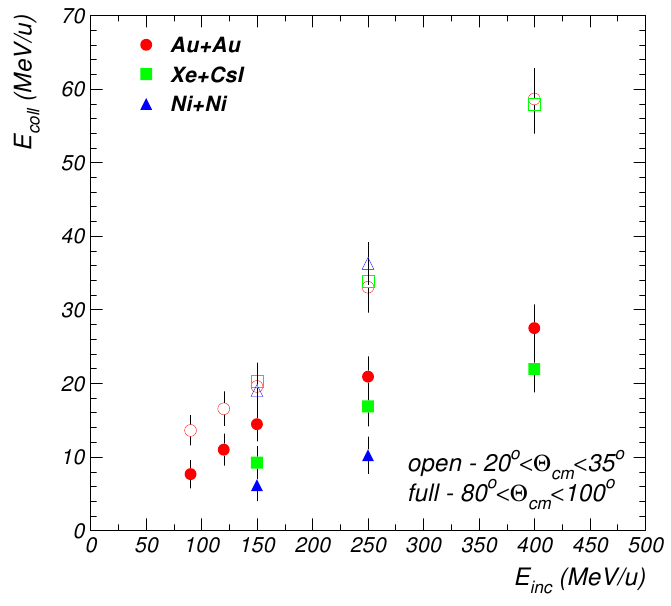}
\caption{Collective energy per nucleon as a function of incident energy 
for three  symmetric systems and highly central collisions (70 mb geometrical cross section) 
for $20^{\circ}<\theta_{cm}<35^{\circ}$ (open symbols) 
and $80^{\circ}<\theta_{cm}<100^{\circ}$ (full symbols) \cite{petro1}.}
\label{fig:Fig2}
\end{figure}

   Detailed studies of collective type expansion of the fireball formed in 
heavy ion collisions as a function of the mass of the colliding nuclei and polar 
angle for highly central collisions \cite{petro1,petro2} evidenced a systematic 
enhancement of collective expansion going from 90 A$\cdot$MeV to 400 A$\cdot$MeV
incident lab energy and from Ni+Ni to Au+Au. 
As can be followed in Fig.~\ref{fig:Fig2}, this conclusion is 
based on the analysis performed around $90^{\circ}$ polar angles in the center of mass system. 
In the forward direction, where Corona effects and fluctuations in the selected 
collision geometry play an important role and the average number 
of collisions suffered by the nucleons emitted in that direction is reduced, 
the extracted $"E_{coll}"$ is artificially larger and its dependence 
on the mass of the colliding system is washed out.   

Under the assumption of a linear dependence of the expansion velocity as a function of the 
position in the fireball, which seems to be a 
rather good approximation at SIS18 energies \cite{petro3}, the average transverse expansion velocity  
can be obtained from the average collective energy per nucleon.
At higher energies, the average transverse expansion velocity and kinetic freeze-out temperature 
were obtained as parameters of expressions 
inspired by hydrodynamical models \cite{heinz} used to fit the experimental transverse 
momentum spectra of different reaction products:
\begin{equation}
E\frac{d^3N}{dp^3}\sim
f(p_t) = \int_{0}^Rm_TK_1(m_Tcosh\rho/T_{fo})I_0(p_Tsinh\rho/T_{fo})rdr
\end{equation} 
where $m_T=\sqrt{m^2+p_T^2}$; $\beta_r(r)=\beta_s(\frac{r}{R})^n$; $\rho=tanh^{-1}\beta_r$.
\begin{figure}[ht]                          
\includegraphics[height=0.3\textheight]{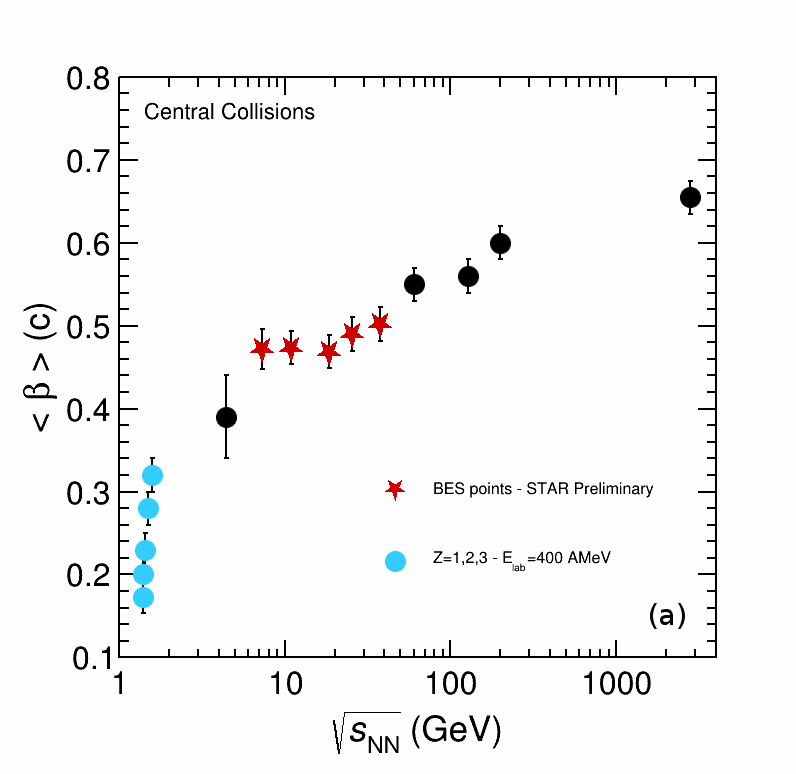}
\includegraphics[height=0.3\textheight]{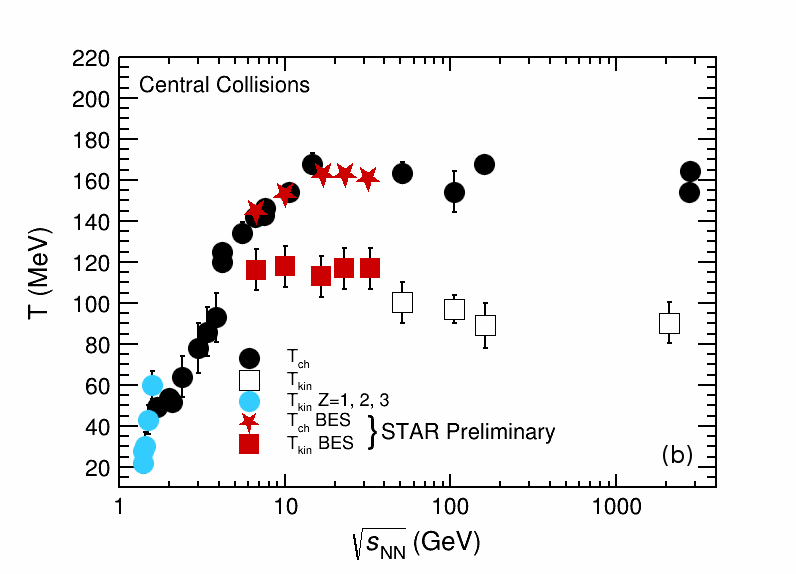}
\caption{(a) $<\beta_T>$ as a function of $\sqrt{s_{NN}}$ for heavy ion central 
collisions \cite{petro2,starscan,E866,star_high,alicePbPb}; 
(b) Chemical ($T_{ch}$) and kinetic freeze-out temperatures for heavy-ion central collisions as a function of the 
collision energy $\sqrt{s_{NN}}$ \cite{starscan,E866,star_high,alicePbPb,stoicea}.}
\label{fig:Fig3}                            
\end{figure}
A compilation in terms of the average transverse expansion velocity ($<\beta_T>$) and the 
kinetic freeze-out temperature (T$_{kin}$) extracted using the above mentioned procedures and of the
chemical freeze-out temperature extracted from fits of yield ratios using 
statistical model expressions for heavy ion central collisions 
can be followed in Fig.~\ref{fig:Fig3}. 
Similar representations could be done as a function of baryon chemical potential $\mu_B$~\cite{starscan}.
As can be seen in Fig.~\ref{fig:Fig3}, the average transverse expansion velocity and kinetic freeze-out temperature 
increase steeply to $\sim$ 50\% of the speed of light and 120 MeV, respectively, 
up to $\sqrt{s_{NN}}\sim$ 8~GeV ($\mu_B$ = 400 - 500 MeV)  followed  
by a  plateau up to \mbox{$\sqrt{s_{NN}}\sim$ 20  GeV} ($\mu_B\sim$ 200 MeV) 
and 40~GeV ($\mu_B\sim$ 100~MeV), respectively. 
At higher collision energies a smooth logarithmic increase of 
the average transverse expansion velocity and decrease of freeze-out temperature up to the LHC energy is evidenced, 
the expectations for LHC based on the extrapolation of the results obtained at lower energies being confirmed.  
The chemical freeze-out temperature increases up to $\sqrt{s_{NN}}\sim$20 GeV and 
remains rather constant at a value of $\cong$160 - 165 MeV.
As far as concerns the dependence of the expansion as a function of centrality, 
starting from $N_{part}\sim$100, $<\beta_T>$ shows a 
very weak dependence on $N_{part}$ starting from 400 MeV up to the LHC energy. 
For mid-central collisions at low energies \cite{stoiceaprl} and also at 200 GeV \cite{shimo} 
it was evidenced that for 
the same $N_{part}$  the more rounded objects expand more violently than the almond shape 
ones if the analysis is done at $90^{\circ}$ relative to the reaction plane. 
In-plane, the results are quite different. While at 200 GeV there is no difference between the 
expansion of fireballs 
corresponding to different symmetric colliding systems for a given $N_{part}$ \cite{shimo}, 
at much lower energies where there is 
a balance between the expansion time and passing time of the spectators and the 
Lorentz contraction is rather negligible, 
the shadowing effects of the spectators play an important role and the observed in-plane expansion 
is lower for 
the heavier symmetric combination for a given $N_{part}$ \cite{stoiceaprl}.      

\section{3. Collision geometry in A-A, \MakeLowercase{p}-A and \MakeLowercase{pp} systems}

   The amount of energy density of the system produced in heavy ion collisions depends on the amount of overlapping matter, therefore on the impact parameter. 
From the fits of the experimental charged particle multiplicity distributions based on Glauber Monte Carlo estimates of the number of ancestors 
and negative binomial distributions for charged particle multiplicity in nucleon-nucleon interactions \cite{miller}, a rather good 
impact parameter selection can be achieved in A-A collisions. 
The situation becomes delicate for p-A collisions where the correlation between the number of wounded nucleons and the impact 
parameter becomes rather broad relative to A-A. 
What about the pp case? 

\begin{figure}[ht]  
\includegraphics[height=0.24\textheight]{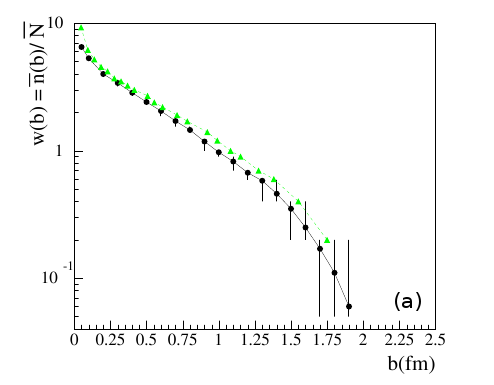}
\includegraphics[height=0.24\textheight]{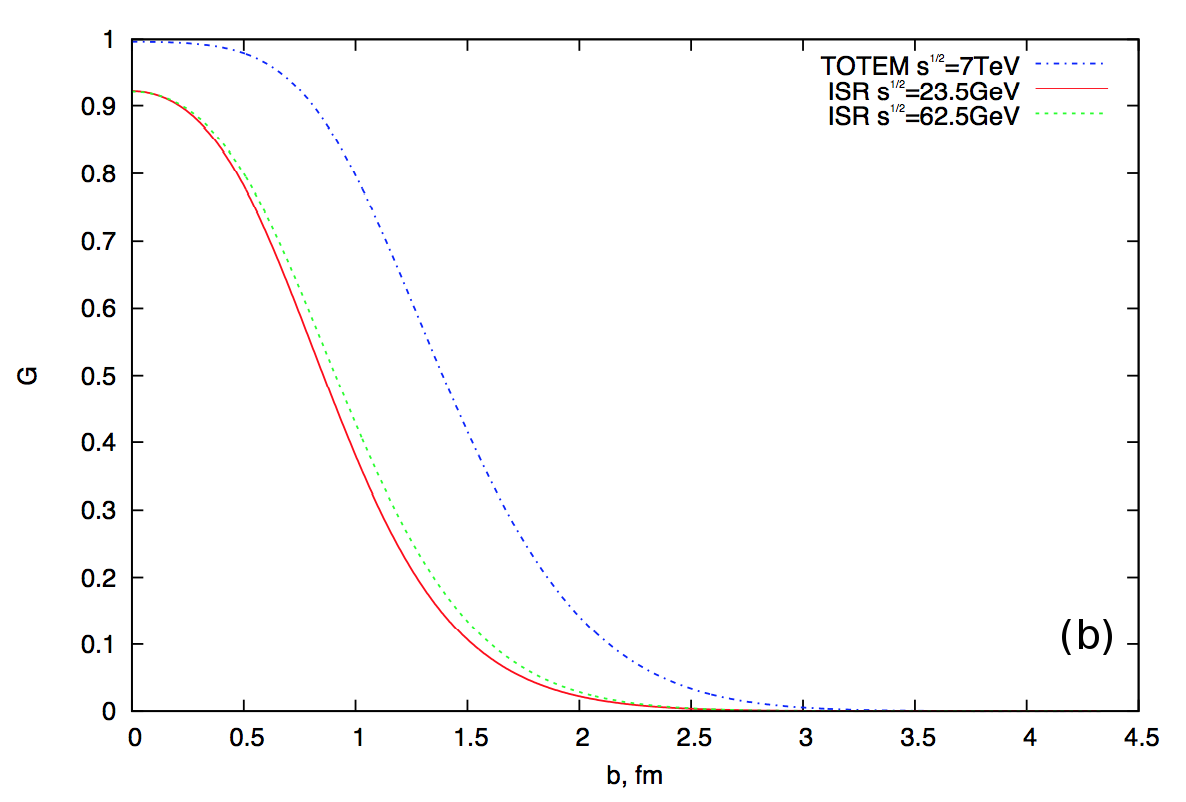}   
\caption{(a) Charged particle multiplicity - impact parameter correlation based on: PYTHIA-like overlap function (black symbols) and 
overlap function from \cite{drem} (green symbols); (b) Overlap function: PYTHIA-like and from \cite{drem}.} 
\label{fig:Fig4}                             
\end{figure}

\begin{figure}[ht]
\includegraphics[height=0.26\textheight]{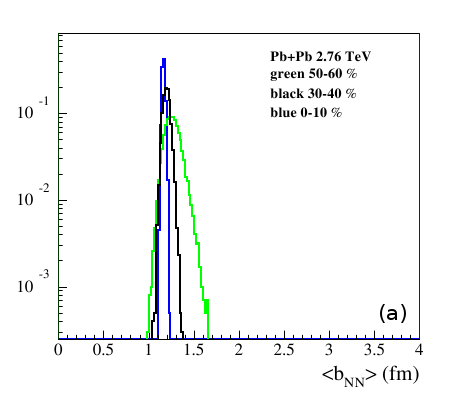}
\includegraphics[height=0.26\textheight]{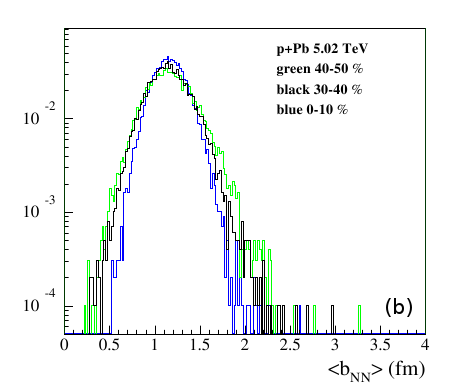}
\caption{(a) The average distance between colliding nucleons at different centralities for Pb-Pb at 2.76~TeV using 3$\cdot10^4$ events; 
(b) The average distance between colliding nucleons at different centralities for p-Pb at 5.02 TeV using 3$\cdot10^4$events.}                           
\label{fig:Fig5}                                  
\end{figure}

Within the geometrical model approach of particle production \cite{moreno, bialas} and 
$\sigma_{in}$(b,s)=1-exp(-k{\it O}(b)) where {\it O}(b) is the overlap function resulting from  spherically symmetric double 
Gaussian distributions of the hadronic matter inside the colliding protons, like in PYTHIA, with parameters fixed in order to get the 
best agreement with the experiment \cite{atlaspythia}, the charged particle multiplicity - impact parameter correlation is obtained by solving the following integral equation:
\begin{equation}
\int_0^{w(b)}\psi(w)dw=\frac{1}{\sigma} \int_b^{\infty}d^2b\sigma(b)
\end{equation}  
where w(b)=$\overline{n}(b)/\overline{N}$, $\overline{N}$P(n)=$\psi(z,\overline{N})$, z=n/$\overline{N}$.  
In order to obtain $\psi(w)$, the charged particle multiplicity distribution \cite{cmsmult} represented in the KNO variable $z=N_{ch}/\overline{N}^{MB}$  was used. 
The result is presented in Fig.~\ref{fig:Fig4}(a) with black points. 
As a cross-check, the overlap function obtained by a recent 
parametrization \cite{drem} (Fig.~\ref{fig:Fig4}(b)) which reproduces 
the elastic differential cross section of pp scattering at 7~TeV \cite{totem} was also used. 
The result is represented in Fig.~\ref{fig:Fig4}(a) with green symbols. 
These results show that at high charged particle multiplicities, i.e. 5-6 times larger than the mean value 
corresponding to the minimum bias multiplicity distribution, events close to central pp collisions could be selected. 
Using the Glauber Monte Carlo approach where the black disk approximation for nucleon-nucleon cross section 
is replaced by the above ansatz for $\sigma_{in}$(b,s) the average distance between colliding nucleons at 
different centralities for Pb-Pb at 2.76~TeV and p-Pb at 5.02~TeV, measured at LHC, can be estimated.
The results for 3$\cdot10^4$ events for each of the two systems are presented in Fig.~\ref{fig:Fig5}.   
The distributions are normalized to the number of events.   

In Pb-Pb collisions at all three centralities represented in Fig.~\ref{fig:Fig5}(a), 
   the percentage of events for which the average distance between the wounded nucleons is smaller than 1.1~fm is rather low. 
In the p-Pb case (Fig.~\ref{fig:Fig5}(b)) there are events in which the average distance goes down to 0.5~fm. 
However, there is no possibility to select such events based on charged particle multiplicity, 
the interplay between the collision geometry and the nucleon-nucleon impact parameter contributions to the charged particle multiplicity being rather complex. 
Therefore, the only way to study phenomena as a function of impact parameter at the nucleonic level is the analysis of pp collisionss as a function of 
charged particle multiplicity and any direct comparison between the three systems in terms of charged particle multiplicity has to be carefully considered.
For a large overlap in a pp collision at high energy, the probability of multi-interactions among the quasi-real cascades of partons and rescattering of the 
struck cascades increases and could contribute to a large energy transfer in a small collision volume of proton size. 
The energy density easily reaches the deconfinement value and considering the mean free path of the constituents of the deconfined phase, thermalization cannot be excluded.
If this is the case, such an initial state, thermalized and of extremely high energy density, will follow a collective 
type expansion \cite{bele,male,vanhov, ander} which can be evidenced
experimentally by analyzing the same observables as the ones presented above for A-A collisions.

\section{4.Transverse momentum spectra of identified charged hadrons and their
ratios in \MakeLowercase{pp} collisions at $\sqrt{s}$ = 7 T\MakeLowercase{e}V 
as a function of charged particle multiplicity} 

\begin{figure}[ht]
\includegraphics[height=0.30\textheight]{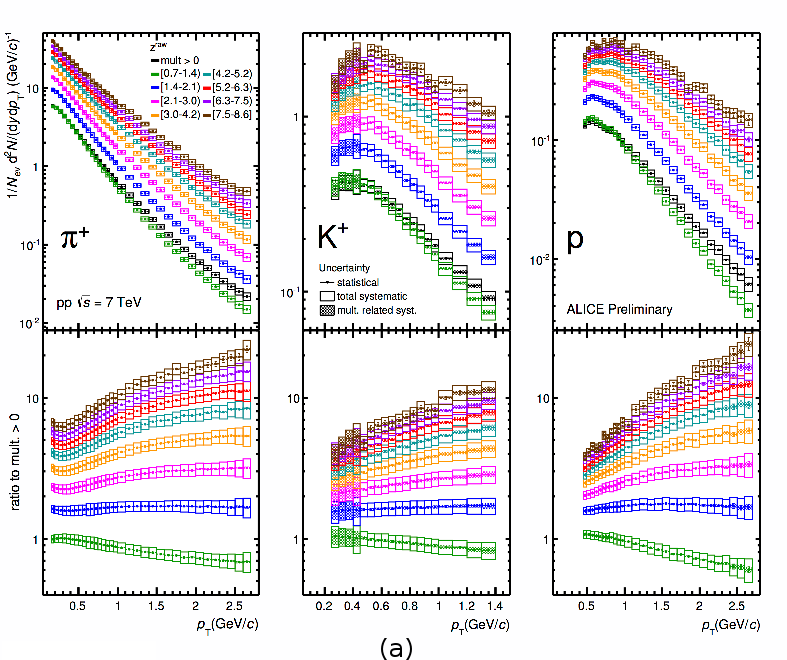}
\includegraphics[height=0.30\textheight]{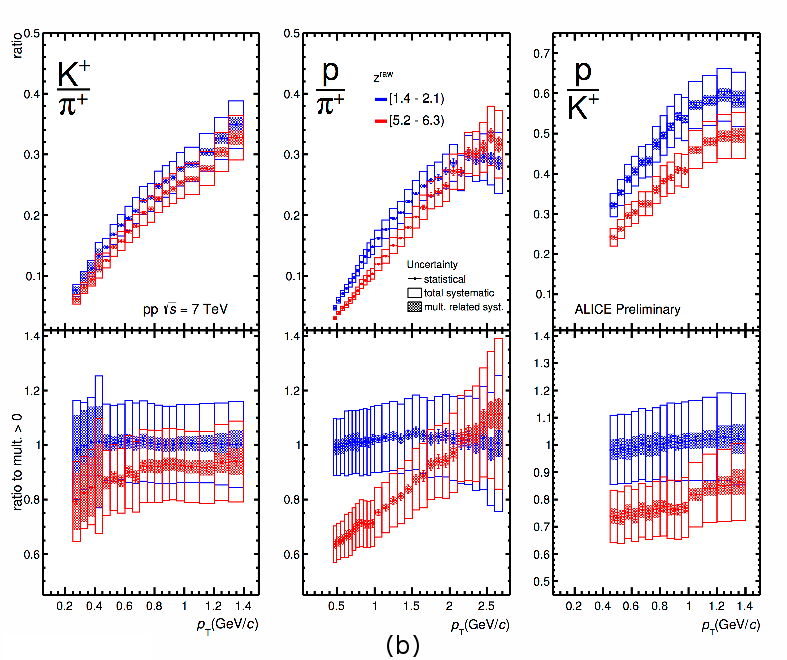}
\caption{(a) $p_T$ spectra and  their ratio to the corresponding MB spectra, 
for $\pi^+$, $K^+$ and p
for eight charged particle multiplicity bins for pp collisions 
at $\sqrt{s}$ = 7 TeV measured by ALICE at LHC. 
(b) $p_{T}$ dependent yield ratios and their ratios to the MB case for two
multiplicity bins \cite{cristi}.}
\label{fig:Fig6}
\end{figure}

\begin{figure}                              
\includegraphics*[height=0.385\linewidth]{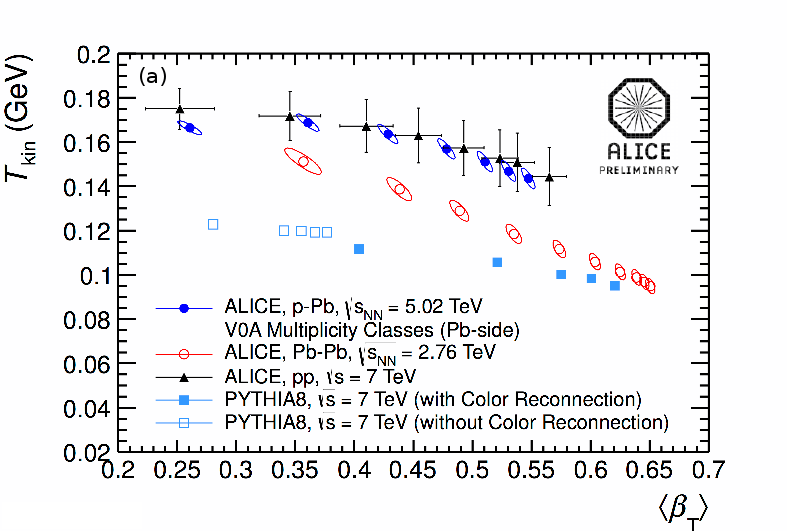}
\includegraphics*[height=0.36\linewidth]{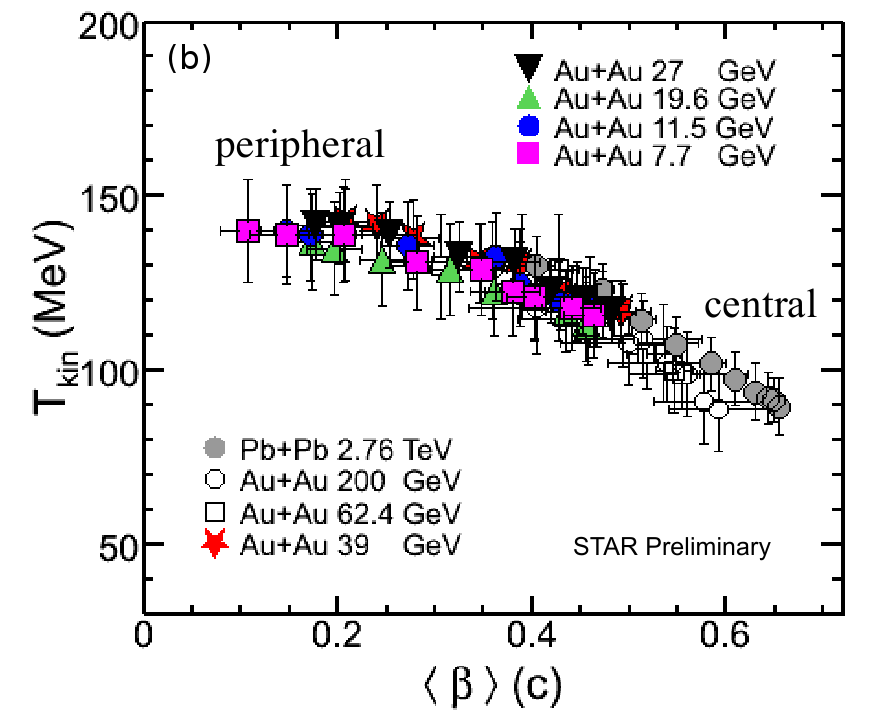}
\caption{(a) $T_{kin}$ - $<\beta_T>$ correlation as a function of multiplicity, multiplicity classes  and 
centrality for pp ($\sqrt{s}$ = 7 TeV), p-Pb ($\sqrt{s_{NN}}$ = 5.02 TeV) 
and Pb-Pb ($\sqrt{s_{NN}}$ = 2.76 TeV), respectively \cite{cristi,alicepPb};  
(b) $T_{kin}$ - $<\beta_T>$ correlation as a function of centrality 
for Au+Au at RHIC energies \cite{starscan} and Pb-Pb \cite{alicePbPb}.}  
\label{fig:Fig7}                            
\end{figure}

   The ALICE Collaboration has recently presented detailed results on transverse momentum spectra of $\pi^+$, $K^+$ and $p$ measured at LHC in pp collisions at $\sqrt{s}$ = 7 TeV 
   as a function of charged particle multiplicity \cite{cristi}.
The charged particle multiplicity was measured in the central pseudorapidity region $|\eta|\le$0.8 and the analysis was done in a narrower range of rapidity $|y|\le$0.5.
The $p_T$ spectra (Fig. 6(a) - upper row) were analyzed from 0.2 GeV/c, 0.3 GeV/c and 0.5 GeV/c up to 2.6 GeV/c, 1.4 GeV/c and 2.6 GeV/c for  $\pi^+$, $K^+$ and p 
respectively and in eight bins of multiplicity up to $\sim$ 50 measured charged particle multiplicity density per unit of pseudorapidity. 
A clear depletion in the 
low $p_T$ region of the spectra ratios to MB as a function of charged particle multiplicity can be seen in  Fig.~\ref{fig:Fig6}(a) - bottom row. 
The ratio of the $p_T$ spectra of $K^+$ and protons relative to $\pi^+$ and protons relative to $K^+$ for the second and the 
sixth multiplicity bins are represented in the upper row and their ratios to the values corresponding to MB case in the bottom row of Fig.~\ref{fig:Fig6}(b). 
At large multiplicities heavier particles, i.e. protons relative to $K^+$ and 
$\pi^+$ and $K^+$ relative to $\pi^+$, are pushed towards large $p_T$ values, in other words depleted in the low $p_T$ range. 
Such a trend, evidenced in Pb-Pb collisions at 2.76 TeV was explained by the existence of a common boost due to
collective expansion \cite{alicePbPb}. 
Qualitatively, a similar trend is observed in the EPOS-LHC model in which a collective hadronization process is included in the pp scattering \cite{eposlhc}.    
This trend was also evidenced recently in p-Pb collisions at 5.02~TeV for central collisions~\cite{alicepPb}. 
As it was presented in the first section, information on collective type dynamics 
in heavy ion collisions from the fits of experimental transverse momentum
spectra using expressions inspired by hydrodynamical models \cite{heinz} was obtained. 
The result of these fits done simultaneously on $\pi^+$, $K^+$ and p spectra, in terms of $T_{kin}$ - $<\beta_T>$ correlation as a 
function of charged particle multiplicity and comparison with the results obtained for  Pb-Pb and p-Pb as a function of 
centrality and multiplicity classes, respectively,  are presented in Fig.~\ref{fig:Fig7}(a) \cite{cristi}.
One could conclude that for pp collisions at 7 TeV, the $T_{kin}$ - $<\beta_T>$ correlation as
a function of charged particle multiplicity 
has a trend rather similar with the one observed in heavy ion collisions, i.e. the freeze-out kinetic 
temperature decreases and the average transverse expansion velocity increases with charged particle multiplicity (pp) or increasing centrality (A-A).
However, there is a quantitative difference between pp and A-A collisions, i.e. $T_{kin}$ is systematically lower and $<\beta_T>$ systematically 
larger than the pp values, the difference increasing towards higher centralities. 
Within the error bars, the results for p-Pb at 5.02 TeV are the same with the ones 
evidenced in pp. 
Such a correlation  is not reproduced by PYTHIA for the pp case. 
Including the color reconnection mechanism \cite{pythiacr} it seems that the model starts to show a similar trend but with values of $T_{kin}$ about 40 MeV lower. 
On Fig.~\ref{fig:Fig7}(b)  a similar plot \cite{starscan} is shown, including all energies measured at RHIC \cite{starscan} and Pb-Pb at 2.76 TeV \cite{alicePbPb}.
\begin{figure}[ht]
\centering
\includegraphics[width=0.308\textheight]{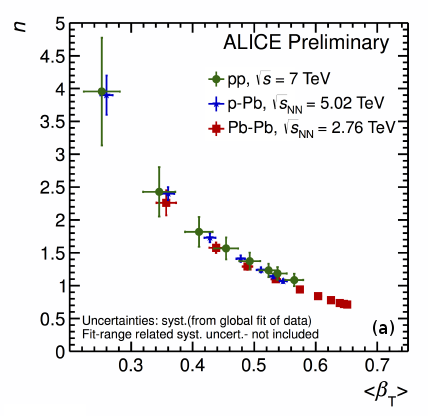}
\includegraphics[width=0.355\textheight]{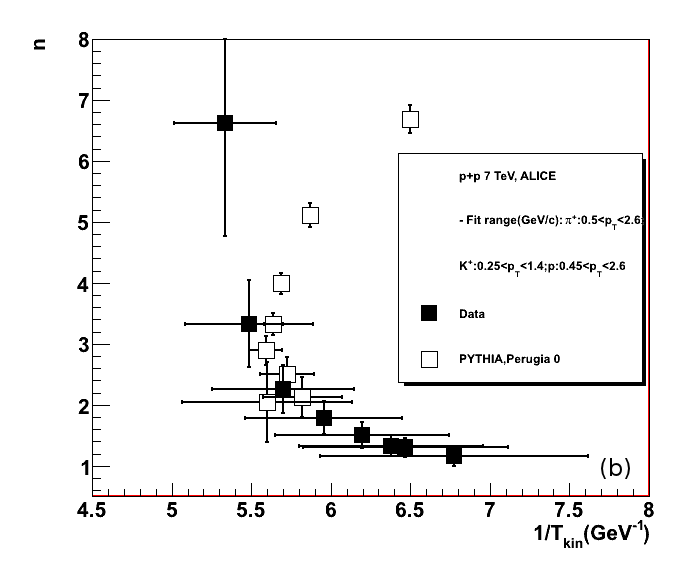}
\caption{(a) n - $<\beta_T>$ correlation as a function of charged particle multiplicity in pp 
at 7 TeV and centrality and multiplicity classes in  Pb-Pb at 2.76 TeV and p-Pb at 5.02 TeV, respectively  \cite{cristi}; 
(b) n - 1/$T_{kin}$ correlation as a function of charged particle multiplicity in pp at 7 TeV; data - full symbols, PYTHIA - open symbols.}  
\label{fig:Fig8}                            
\end{figure}

Another aspect worth to be mentioned is the correlation between the expansion profile (n) and $<\beta_T>$. 
This correlation is presented in Fig.~\ref{fig:Fig8}(a) \cite{cristi}. 
It is clearly seen that all three systems follow exactly the same correlation.
Towards the highest multiplicity in the pp case, the expansion velocity becomes linear as a function of position within the fireball.
The n - 1/T$_{kin}$ correlation (Fig.~\ref{fig:Fig8}(b)) obtained using PYTHIA results is completely different than the experimental one.  
Using  EPOS3 estimates of the mean transverse momentum ($<p_T>$) for $\pi^+$, $K^+$ and p as a function of multiplicity \cite{epos3}, the relative yields 
measured by the CMS Collaboration in pp collisions at 7 TeV \cite{cmsidchpart_mult} and the impact parameter-charged particle 
multiplicity correlation presented in Fig.~\ref{fig:Fig4}(a), a rough estimate of the Bjorken energy density 
times the formation time, i.e. $\varepsilon_{Bj}\cdot\tau$, was made for the highest charged particle 
multiplicity measured by CMS within the  pseudorapidity range $|\eta|\leq2.4$. The obtained value is $\simeq$ 10 GeV/$fm^2$. 
Based on PYTHIA simulations, the largest multiplicity bin measured by CMS would correspond to the fifth multiplicity bin measured by ALICE~\cite{cristi}.
Therefore, quite probable at the largest multiplicity bin measured by ALICE, 
the Bjorken energy density is similar with the value of $\sim$17.5 GeV/fm$^2$ estimated for the highest centrality (0-5\%) Pb-Pb collisions at 2.76 TeV.  
\begin{figure}[ht]
\includegraphics[height=0.17\textheight]{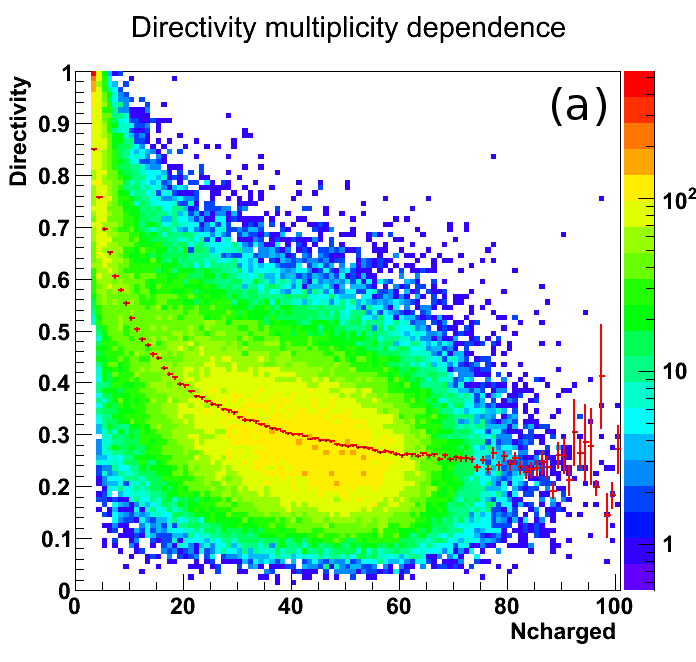}
\includegraphics[height=0.17\textheight]{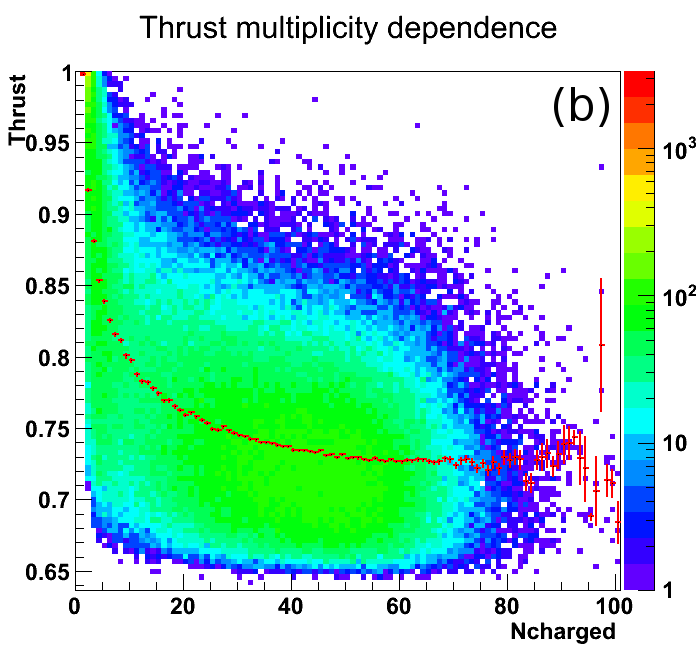}
\includegraphics[height=0.17\textheight]{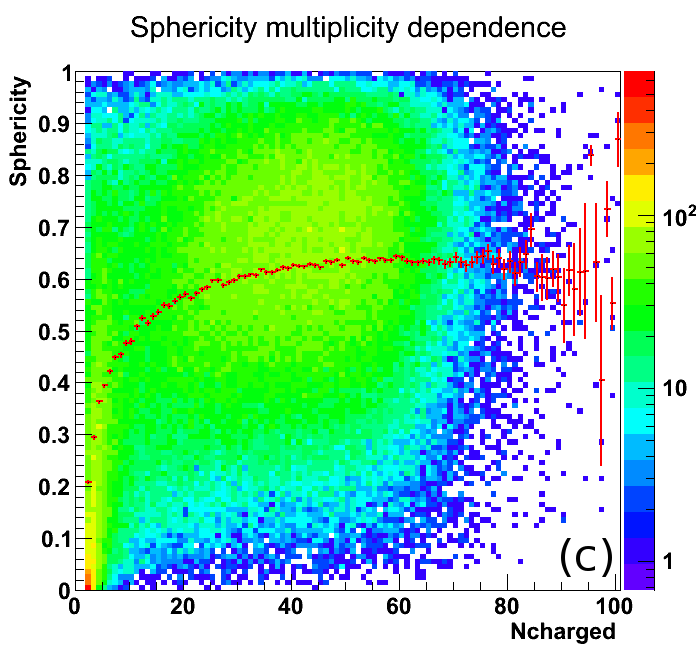}
\includegraphics[height=0.17\textheight]{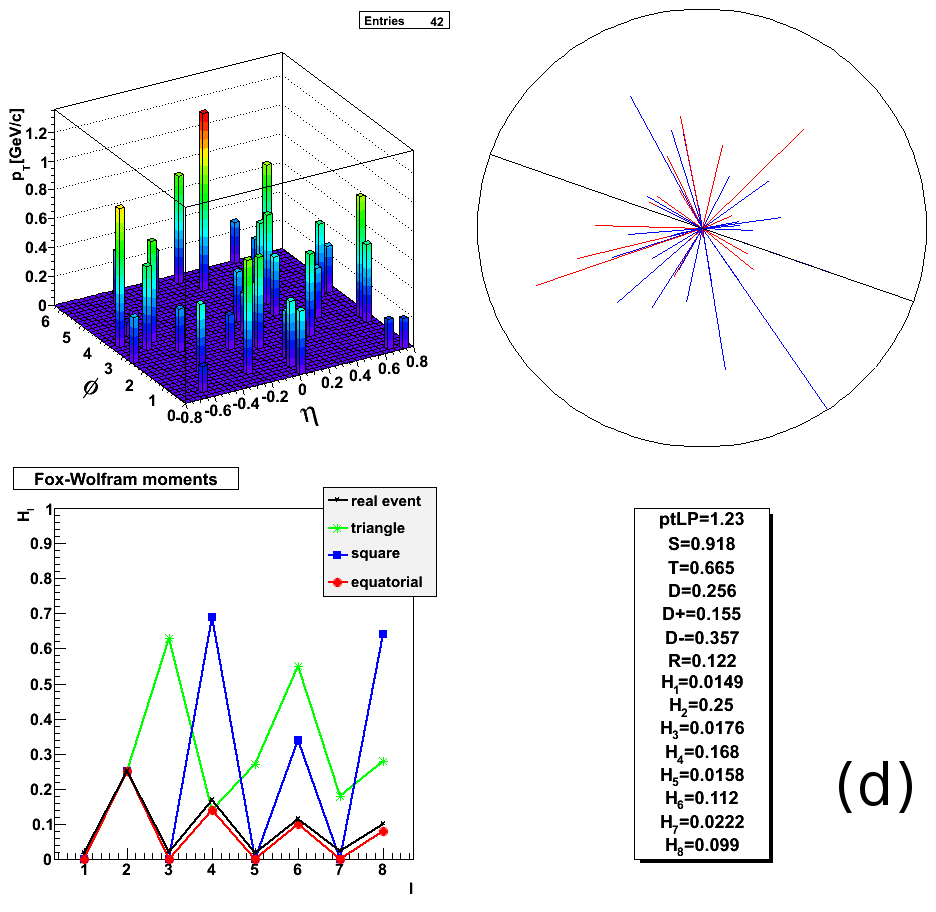}
\caption{(a) Directivity - multiplicity correlation for PYTHIA pp collisions at 7 TeV \cite{cristiphd, andreiphd}; 
(b) Thrust - multiplicity correlation for PYTHIA pp collisions at 7 TeV \cite{andreiphd}; 
(c) Sphericity - multiplicity correlation for PYTHIA pp collisions at 7 TeV \cite{andreiphd}; 
(d) An example of selection performace of an azimuthal isotropic event using Fox-Wolfram moments \cite{andreiphd}.}  
\label{fig:Fig9}
\end{figure}

Although these preliminary results seem to indicate that for  events with high charged particle multiplicity  
  in pp collisions at 7~TeV collective type phenomena show up, the experimental and theoretical efforts to understand the physics behind the evidenced trends remain to be done. 
On the experimental side one important aspect is to discriminate between hard and soft processes, jets influence on the analysis based only on multiplicity selection 
not being negligible. 
Studies along the possibility to select events close to azimuthal isotropy using global event shape 
observables like Directivity, Sphericity, Thrust or Fox-Wolfram moments have shown their performance in 
selecting soft, nearly azimuthal isotropic events \cite{cristiphd,andreiphd}. 
As an example, the correlations between 
charged particle multiplicity and Directivity, Thrust
and Sphericity, based on PYTHIA events for pp at 7~TeV, are shown in Fig.~\ref{fig:Fig9}. 
Although the correlations are rather good, it is clearly seen that even at the highest multiplicities the global event shape variables have a rather broad distribution. 
Therefore, a two dimensional condition in multiplicity and different event shape variables could significantly contribute in 
selecting events with specific azimuthal distributions for a given multiplicity. 
An example in this respect is presented in Fig.~\ref{fig:Fig9}(d) where an azimuthal isotropic event was selected based on \mbox{Fox-Wolfram} moments. 
The two-dimensional distribution of particles for such an event in $\eta$-$\phi$ and $p_x$-$p_y$ confirms the expectation.     
\begin{figure}[ht]
\includegraphics[height=0.26\textheight]{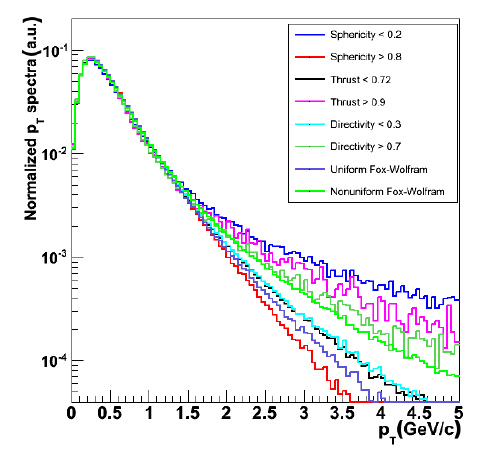}
\caption{PYTHIA $p_T$ spectra of positive charged particles for MB events for jet-like and close to azimuthal 
isotropic events selected by different event shape global variables \cite{andreiphd}.}
\label{fig:Fig10}
\end{figure}

An example of the influence of such selections on the $p_T$ spectra of charged particles obtained with PYTHIA, is shown in Fig.~\ref{fig:Fig10}. 
In general, selecting events close to isotropy based on the four global event shape variables reduces significantly the power law tail of 
the $p_T$ spectra.

\section{5. Conclusions}
Evidence for collective type expansion of the hot and compressed fireball produced in heavy ion collisions is presented for all collision energies, 
starting from below $\sqrt{s_{NN}}$ = 1.5~GeV up to 2.76~TeV. 
The expansion velocity increases steeply to about half of the speed of light up to $\sqrt{s_{NN}}$ = 8 GeV, corresponding to a 
baryonic chemical potential where the thermal degrees of freedom start to be dominated by mesons. 
Above this value, the expansion velocity shows a plateau up to $\sqrt{s_{NN}}\sim$ 20~GeV when it starts to increase 
again logarithmically with a much lower slope up to the LHC energy. 
The kinetic freeze-out temperature follows a similar trend up to $\sqrt{s_{NN}}\sim$ 40~GeV beyond which it slowly decreases, the system being cooled down during the expansion.  
The chemical freeze-out temperature increases up to $\sqrt{s_{NN}}\sim$ 20 GeV
and remains rather constant at a value of $\sim$ 160 - 165~MeV.
At a given number of participating nucleons, the almond shape fireballs expand less violently along the major axis than the round shape ones.
Above $N_{part}\sim$ 100, $<\beta_T>$ shows a very weak dependence on $N_{part}$ for all collision energies.

Transverse momentum distributions and their ratios for $\pi^+$, $K^+$ and p at mid-rapidity (|y| <0.5) for different 
charged particle multiplicity bins in pp collisions at $\sqrt{s}$ = 7~TeV show an enhanced depletion of heavier species 
relative to the lighter ones in the low $p_T$ region with increasing charged particle multiplicity. 
The quality of simultaneous fits of the experimental spectra using the Boltzmann-Gibbs Blast Wave  expression and the 
dynamics of the extracted kinetic freeze out temperature, average transverse expansion velocity  and its profile  as a function of 
multiplicity are similar with those obtained in heavy ion collisions.
Preliminary estimates of the Bjorken energy density for high multiplicity events
indicate values close to the ones estimated for the central Pb-Pb collisions at
$\sqrt{s_{NN}}$ = 2.76 TeV. 
The selection of high multiplicity events close to azimuthal isotropy based on event shape global observables seems to be feasible.
A direct comparison among pp, p-Pb and Pb-Pb based on charged particle multiplicity has to be taken with care.
A factor two in the collision energy, soon available at LHC and the expected higher statistics 
will give access to extend these studies at heavy flavor hadrons and detailed comparisons  with the results obtained in A-A collisions.

\bibliography{mihai_petrovici_et_al_bibnou}{}

\end{document}